\documentstyle[aps,prl,epsfig,twocolumn]{revtex} 

\begin{document}
\pagestyle{myheadings} 
\markboth{Huberman/Helbing: Optimizing Traffic Flow}
{Huberman/Helbing: Optimizing Traffic Flow}
\tighten
\onecolumn
\twocolumn[\hsize\textwidth\columnwidth\hsize\csname @twocolumnfalse\endcsname

\title{Optimizing Traffic Flow}
\author{Bernardo A. Huberman$^{+}$ and Dirk Helbing$^{\ast }$}
\address{$^{+}$ Xerox PARC, 3333 Coyote Hill Road, Palo Alto, CA 94304, USA\\
$^{\ast }$ II. Institute of Theoretical Physics, University of Stuttgart,
Pfaffenwaldring 57/III, 70550 Stuttgart, Germany}
\maketitle
\draft

\begin{abstract}
We present an economics-based method for deciding the optimal rates at which
vehicles are allowed to enter a highway.  The method exploits the naturally
occuring fluctuations of traffic flow and is flexible enough to adapt in
real time to the transient flow characteristics of road traffic. Simulations
based on realistic parameter values show that this strategy is feasible  for
naturally occurring traffic, and that even far from optimality, injection
policies can improve traffic flow. Our results also allow a better
understanding of the high flows observed in ``synchronized'' congested
traffic close to ramps.
\end{abstract}

]

%

The advent of powerful traffic simulators \cite{NS,zellauto,nature} has led
to a spate of new discoveries in the area of vehicular traffic that agree
well with empirical observations \cite{TGF97}. 
From moving traffic jams \cite{NS} to transitions to states of coherent
motion \cite{nature} these new results offer insights into a complicated and
socially relevant many-body problem, while also suggesting ways of designing
controls that can maximize the flow of vehicles in cities and highways \cite
{Firefly,Bovy}.

The recent discovery of a new state of congested highway traffic, called
``synchronized'' traffic \cite{sync}, has generated a strong interest in the
rich spectrum of phenomena occuring 
close to on-ramps \cite{Lee,HT}. In this connection, a particularly
relevant problem is that of choosing an optimal injection\ strategy of
vehicles into the highway. Similar questions have recently been raised
regarding the most efficient use of the Internet in light of its bursty
congestion patterns\cite{Internet}. While there exist a number of {\em %
heuristic} approaches to optimizing vehicle injection into freeways by
on-ramp controls, the results are still not satisfactory. What
is needed is a strategy that is flexible enough to adapt in real time to the
transient flow characteristics of road traffic while leading to minimal
travel times for all vehicles on the highway. 

This paper presents a solution to this problem that explicitly exploits the
naturally occuring fluctuations of traffic flow in order to enter the
freeway at optimal times. This method, which leads to a more homogeneous
traffic flow and a reduction of inefficient stop-and-go motions, is less
sensitive to failures in the control mechanism than traditional approaches. 

The basic performance criterion behind this optimization technique is the
travel time distribution of vehicles \cite{Chaos} which is to be contrasted
with the local velocity distribution applied in other approaches. The travel
time distribution is a global measure of the overall dynamics on the whole
freeway stretch. It allows the evaluation of both the expected arrival time
of vehicles at a destination and its variance, the latter characterizing the
likelihood that a particular vehicle will have a travel time different from
the expected one. Both the average and the variance of travel times are
influenced by the inflow of vehicles entering the freeway over an on-ramp.
From these two quantities one can construct a relation between 
the average payoff (the negative mean value of travel times) 
and the risk (their variance), as is considered in many optimization problems
in economics. The optimal strategy will then correpond to the point in
the curve that yields the lowest risk at a high average payoff.
In the following, we will show that 
the variance of travel times has a minimum for on-ramp flows
that are different from zero, implying that
traffic flow can be optimized by choosing the appropriate vehicle injection
rate into the freeway.

In order to obtain the travel time distribution of vehicles on a highway, we
simulated two-lane traffic flow via a discretized follow-the leader model,
which describes the empirical known features of traffic flows quite well 
\cite{zellauto}. \ In our experiments, we extended the simulation to several
lanes with lane-changing maneuvers and different vehicle types (cars and
trucks). We determined the travel times of all vehicles by storing the times
at which they pass two successive cross sections of the road. Empirical data
of this kind could be collected at the gates of a toll road. The time
difference between the passing times corresponds to the travel time needed
to traverse the freeway stretch between the two cross sections. In our
simulations, we chose a stretch of length \ $L=10$\thinspace km.

The model distinguishes $I$ neighboring lanes $i\in \{1,\dots ,I\}$ of a
unidirectional freeway. All lanes are subdivided into sites $z\in
\{1,2,\dots ,Z\}$ of equal length $\Delta x=2.5$\thinspace m. Each
site is either empty or occupied, the latter case representing the back of 
a vehicle of type $a\in \{1,\dots ,A\}$ 
with velocity $v = u \Delta x / \Delta t$. Here, 
$u \in \{0,1,\dots, u_a^{\rm max} \}$ is the number of sites that the vehicle
moves per update step $\Delta t$. We have
distinguished cars ($a=1$) and trucks ($a=2$). These are characterized by
different optimal velocities $U_{a}(d)$ with which the vehicles would like
to drive at a distance $d$ to the vehicle in front. Their lengths $l_{a}$
correspond to the maximum distances satisfying $U_{a}(l_{a})=0$. The
positions $z(T)$, velocities $u(T)$, and lanes $i(T)$ of all vehicles are
updated \cite{upd} every time step $\Delta t=1$\thinspace s at times $T\in
\{1,2,\dots \}$ according to the following successive steps \cite{nature}:

1.~{\em Determine the potential velocities} $u_{j}(T+1)$ on the present and
the neighboring lanes $i(T)+j$ with $j\in \{-1,0,+1\}$ according to the {\em %
acceleration law} 
\begin{equation}
u_{j}(T+1)=\big\lfloor\lambda U_{a}\big(d_{j}(T)\big)+(1-\lambda )u(T)%
\big\rfloor\,.
\end{equation}
Here, the floor function $\lfloor x\rfloor $ is defined by the largest
integer $m\leq x$, and $d_{j}(T)=[z_{j}^{+}(T)-z(T)]$ denotes the distance
to the next vehicle ahead (+) on lane $i(T)+j$ at position $z_{j}^{+}(T)$.
The above equation describes the typical follow-the-leader behavior of
driver-vehicle units. Delayed by the reaction time $\Delta t$, they tend to
move with the distance-dependent optimal (safe) velocity $U_{a}$, but the
adaptation takes a certain time $\tau =\lambda \,\Delta t$ because of the
vehicle's inertia. Good results, in the sense of replicating highway data,
are obtained for $\lambda =0.77$.

2.~{\em Change lane} to the left, i.e. set 
\begin{equation}
i(T+1)=i(T)+k
\end{equation}
with $k=+1$, if the  vehicle considered can go fastest there, i.e. if 
\begin{equation}
u_{+1}(T+1)>u_{0}(T+1)\quad \mbox{and}\quad >u_{-1}(T+1)\,.
\end{equation}
This is usually the case if the headway on the left lane is greatest. Apart
from the validity of this {\em incentive criterion,} we demand two extra 
{\em safety criteria} \cite{Two}: First, the current vehicle position should
be ahead of the expected position $z_{+1}^{-}(T+1)$ of the following vehicle
($-$) on the left lane, i.e. 
\begin{equation}
z(T)>z_{+1}^{-}(T+1)\,.
\end{equation}
Second, the potential velocity on the left lane should not be considerably
less than the expected velocity $u_{+1}^{-}(T+1)$ of the following vehicle,
i.e. 
\begin{equation}
u_{k}(T+1)\geq q\,u_{+1}^{-}(T+1)\,.
\end{equation}
Once again, realistic results are obtained for $q=0.7$. A value $q<1$
implies that drivers are ready to accept a braking maneuver of the follower
on the destination lane at the next update step. Therefore, the values of $q$
are a measure of how relentless drivers are in overtaking.

Assuming symmetrical (``American'') lane changing rules for simplicity, a
change to the right lane ($k=-1$) is carried out, if the incentive criterion $%
u_{-1}(T+1)>u_{0}(T+1)$ and $\geq u_{+1}(T+1)$ as well as the safety
criteria $z(T)>z_{-1}^{-}(T+1)$ and $u_{k}(T+1)\geq q\,u_{-1}^{-}(T+1)$ are
fulfilled. Otherwise the vehicle stays on the same lane ($k=0$).

3.~If the potential velocity $u_k(T+1)$ on the new lane $i(T+1)$ is
positive, diminish it by 1 with probability $p=0.001$, which accounts
for {\em delayed adaptation} due to reduced attention of the driver and the
variation of vehicle velocities: 
\begin{equation}
u(T+1) = u_k(T+1) - \left\{ 
\begin{array}{ll}
1 & \mbox{with probability $p$} \\ 
& \mbox{if $u_k(T+1)>0$} \\ 
0 & \mbox{otherwise.}
\end{array}
\right.
\end{equation}
The resulting value defines the updated velocity $u(T+1)$.

4.~Update the vehicle position according to the {\em equation of motion} 
\begin{equation}
z(T+1) = z(T) + u(T+1) \, .
\end{equation}

Our simulations were carried out for a circular two-lane road. After the
overall density was selected, vehicles were homogeneously distributed over
the road at the beginning, with the same densities on both lanes. The
experiments started with uniform distances among the vehicles and their
associated desired velocities. The vehicle type was determined randomly
after specifying the percentages of cars (90\%) and trucks (10\%).

At a given injection rate, vehicles enter the beginning of an on-ramp lane
(i.e. a third lane) of 1 kilometer length with a uniform time headway and
their optimal velocity related to the density in the destination lane. Our
simulation stretch consists of a 1 km long 3-lane part and the $L=10$ km
long two-lane stretch on which the travel times are measured. No vehicle is
allowed to change to the on-ramp from the main road. Injected vehicles try
to change from the on-ramp to the main road as fast as possible, i.e.
according to the above lane-changing rules, but they do not care about the
incentive criterion. The end of the on-ramp is treated
like a resting vehicle, i.e. any vehicle that approaches it has to stop, but
it changes to the destination lane as soon as it finds a sufficiently large
gap. If the on-ramp is completely occupied by vehicles waiting to enter the
main road, the injected vehicles form a queue and enter the on-ramp as soon
as possible. Injected vehicles that have completed their 10 km trip on the
two-lane measurement stretch are automatically removed from the freeway
(which would correspond to uncongested off-ramps adjacent to the lanes).
Since our evaluations started after a transient period of two hours and
continued until the 1000th injected vehicle finished its trip, the results
are largely independent of the initial conditions.

\begin{figure}[tbp]
\begin{center}
\epsfig{height=8\unitlength, angle=-90, 
      bbllx=50pt, bblly=50pt, bburx=554pt, bbury=770pt, 
      file=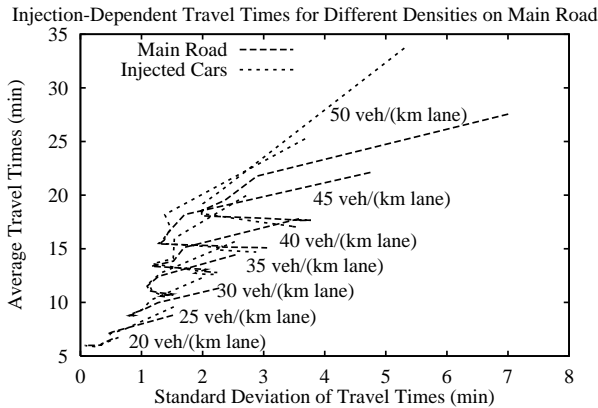} 
\end{center}
\caption[]{Average and standard deviation of the travel times
of vehicles on the main road and from 
the ramp as a function of the injection rate 
for various vehicle densities on the main road (measured without injection). 
With increasing injection rate, the average travel times are increasing
due to the higher resulting vehicle density on the freeway. 
\label{F1}}
\end{figure}

If we plot the average of travel times as a function of their standard
deviation (Fig.~\ref{F1}), we obtain curves parametrized by the injection
rate of vehicles into the road. Several points are worth noting: 1. With
growing injection rate $Q_{{\rm rmp}}=1/2^{n}{\rm s}$ ($n\in \{2,3,\dots
,10\}$), the travel time increases monotonically. This is because of the
increased density caused by injection of vehicles into the freeway. 2. The
average travel time of {\em injected} vehicles is higher, but their
standard deviation lower than for the vehicles circling on the main road.
This is due to the fact that vehicle injection produces a higher density on
the lane adjacent to the on-ramp, which leads to smaller velocities. The
difference between the travel time distributions of injected vehicles and
those on the main road decreases with the length $L$ of the simulated road,
since lane-changes tend to equilibrate densities between lanes. 

In addition, the standard deviation of travel times has a {\em minimum} 
for {\em finite}
injection rates, as entering vehicles tend to fill existing gaps and thus
homogenize traffic flow. This minimum is optimal in the sense that there is
no other value of the injection rate that can produce travel with smaller
variance. In particular, gap-filling behavior mitigates inefficient
stop-and-go traffic at medium densities. Above a density of 45 vehicles per
kilometer and lane on the main road (without injection), the minimum of the
travel times' standard deviation occurs for $n=6$. The reduction of the
average travel time by smaller injection rates is less than the increase
of their standard deviation. This result suggests that, in order to generate
predictable and reliable arrival times, one should operate traffic at medium
injection rates. Lastly, for the case of 40 vehicles per kilometer and lane,
the minimum of the standard deviation of travel times is located at $n=5$,
while for 35 vehicles per kilometer and lane, it is at $n=4$. Below 30
vehicles per kilometer and lane, vehicle injection does not reduce the
standard deviation of travel times. This is because at these densities
homogeneous traffic is stable anyway, so no stop-and-go traffic exists and
therefore no large gaps that can be filled \cite{nature}.

The curves displayed in Fig.~\ref{F1} correspond to a given density $\rho _{%
{\rm main}}$ on the main road {\em without} injection of vehicles. 
The effective
density $\rho _{{\rm eff}}$ on the freeway {\em resulting} from the injection 
of vehicles can be approximated by 
\begin{equation}
\rho _{{\rm eff}}=\rho _{{\rm main}}+\frac{N_{{\rm inj}}}{IL}\,,
\end{equation}
where $I=2$, $L=10$ km. $N_{{\rm inj}}$ is the average number of
injected vehicles present on the main road and can be written as 
\begin{equation}
N_{{\rm inj}}=N_{{\rm tot}}\frac{{\cal T}_{{\rm inj}}}{{\cal T}_{{\rm tot}}-%
{\cal T}_{{\rm inj}}}\,,
\end{equation}
where $N_{{\rm tot}}=1000$ is the total number of injected vehicles during
the simulation runs, ${\cal T}_{{\rm inj}}$ is their average travel time,
and ${\cal T}_{{\rm tot}}$ the time interval needed by all $N_{{\rm tot}}
= 1000$
vehicles to complete their trip. We point out that, in addition to these
measurements, we used two other methods of density measurement which
yielded similar results.

\begin{figure}[tbp]
\begin{center}
\epsfig{height=8\unitlength, angle=-90, 
      bbllx=50pt, bblly=50pt, bburx=554pt, bbury=770pt, 
      file=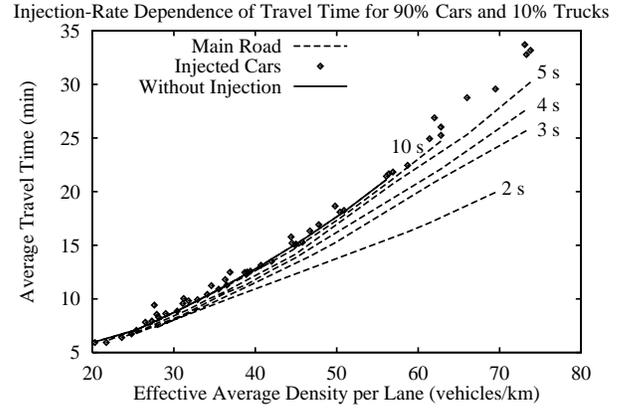} 
\end{center}
\caption[]{Average travel times of vehicles on the main road
as a function of the resulting effective
vehicle density on the freeway, for various injection rates.
In the limit of high injection rates, one observes the predicted linear
dependence of average travel times on effective density, see
Eq.~(\ref{see}). In contrast to the vehicles on the main road, the 
travel times of injected vehicles did not depend on the injection rate.
However, when we checked what happens if the vehicles on the main road 
try to change to the left lane along the on-ramp
in order to give way to entering vehicles,
we found that both, injected
vehicles and the vehicles on the main road, profited. 
\label{F2}}
\end{figure}

We also investigated the dependence of the travel time characteristics on
the resulting {\em effective} densities of vehicles. As Figure~\ref{F2} shows,
vehicle injection can actually reduce the average travel times of the
vehicles on the main road, while the travel times of injected cars are about
the same as those of vehicles on the main road without injection. This means
that, for given $\rho_{\rm eff}$,
one can actually increase the average velocity $V_{{\rm main}}=L/{\cal T%
}_{{\rm main}}$ of vehicles by injecting vehicles at a high rate without
affecting their travel times. This\ result follows from the increased
degree of homogeneity caused by entering vehicles that fill gaps on the main
road, which mitigates the less efficient stop-and-go traffic. It might
actually make sense to diverge a certain proportion of vehicles at
intermittent times and to reinject them later following our methodology.

\begin{figure}
\begin{center}
\epsfig{height=8\unitlength, angle=-90, 
      bbllx=50pt, bblly=50pt, bburx=554pt, bbury=770pt, 
      file=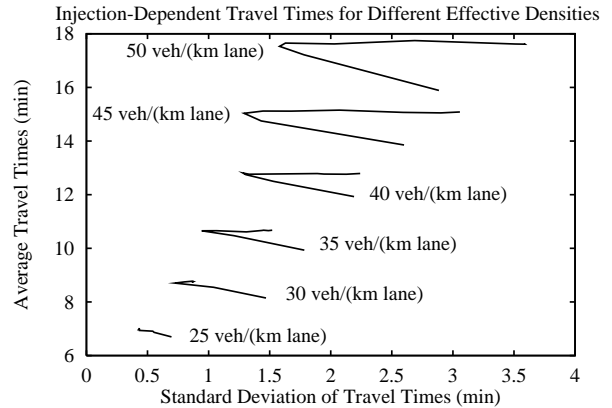} 
\end{center}
\caption[]{As figure~\ref{F1}, but as a function of the 
resulting effective density on the freeway. 
We find shorter travel times at high injection rates
because of the homogenization of traffic. 
The standard deviation of travel
times is varying stronger than the average travel time, which 
indicates that medium injection rates are the optimal choice at high
vehicle densities. 
\label{F3}}
\end{figure}

Finally, Figure~\ref{F3} shows the average of 
the travel times for vehicles in the
main road as a function of their standard deviation. In contrast to Fig.~\ref
{F1}, the curves were computed for the resulting 
{\em effective} densities. This
time, an increase of the injection rate (which corresponds to a smaller
number of circling vehicles on the main road and a greater proportion of
injected vehicles) {\em reduces} the average travel times. Once again, we
observe a minimum of the standard deviation of travel times at high vehicle
densities and medium injection rates.

In the limit of high injection rates, a traffic jam of maximum density $\rho
_{{\rm max}}$ builds up at the end of the on-ramp, while downstream of it we
find the typical density $\rho _{{\rm out}}$ related to the universal
outflow $Q_{{\rm out}}$ from traffic jams \cite{charak,zellauto}. 
We conjecture that the resulting
structure consists of a block of density $\rho _{{\rm max}}$ and length $%
L_{1}$ containing $N_{1}=\rho _{{\rm max}}L_{1}$ vehicles, and a block of
density $\rho _{{\rm out}}$ of length $(L-L_{1})$ containing $(N-N_{1})=\rho
_{{\rm out}}(L-L_{1})$ vehicles at a mean density of 
$\rho_{\rm eff}=N/L$.
The expected travel time ${\cal T}_{{\rm main}}$ would be 
\begin{eqnarray}
{\cal T}_{{\rm main}} &=&\frac{L}{V_{{\rm out}}}+\frac{L_{1}}{C}
 + {\cal T}_{{\rm acc}}+{\cal T}_{{\rm dec}}  \nonumber \\
&=&\frac{L}{V_{{\rm out}}}+\frac{\rho _{{\rm eff}}-\rho _{{\rm out}}}{\rho _{%
{\rm max}}-\rho _{{\rm out}}}\frac{L}{C}+ {\cal T}_{{\rm acc}}+{\cal T}_{%
{\rm dec}} \,,
\label{see}
\end{eqnarray}
where $V_{{\rm out}}$ is the typical velocity emerging downstream of a
traffic jam, and $C$ is the universal dissolution velocity of traffic jams 
\cite{charak,zellauto}. 
Notice that, for high injection rates, the average
travel time should grow {\em linearly} with the mean density $\rho _{{\rm eff%
}}$, which is consistent with the results displayed in Fig.~\ref{F2}. For
decreasing injection rates, travel times should increase, since the
alternation of congested and free flow in the resulting stop-and-go traffic
implies relevant acceleration times ${\cal T}_{{\rm acc}}$ and deceleration
times ${\cal T}_{{\rm dec}}$ in total.

In conclusion, we have presented a strategy for optimizing
traffic on highways in the sense of higher flows and
more reliable predictions of individual travel times.
The applied method is economics-based and
resorts to the establishment of average
payoff versus risk curves. Here, the average payoff corresponds to the
negative mean value of travel times and the risk to their variance. 
The strategy exploits the naturally occuring fluctuations of traffic
flow in order to allow \ the entry of new vehicles to the freeway at optimal
times. Simulations based on realistic parameter values show that this
strategy is feasible for naturally occurring traffic, and that even far from
optimality, injection policies can improve traffic flow. The latter 
allows an interpretation of the high flows observed in ``synchronized''
congested traffic close to ramps \cite{sync}, 
in contrast to the lower average flows
in stop-and-go traffic.

{\em Acknowledgments:} D.H. wants to thank the DFG for financial support
(Heisenberg scholarship He 2789/1-1).

\begin{references}
\bibitem{upd}  
We applied sequential update in driving direction, which avoids conflicts of
vehicles that like to change to the same lane and position from both
neighboring lanes at the same time. For two-lane roads the sequential update
leads to almost identical results as a parallel update, which is usually
applied in cellular or lattice gas automata for flow simulations and gives
realistic results.
\end{references}
\end{document}